\documentclass[11pt,fleqn]{article}
\usepackage{amsmath,amssymb,graphicx}

\begin{document}
\sloppy
\newtheorem{axiom}{Axiom}[section]
\newtheorem{conjecture}[axiom]{Conjecture}
\newtheorem{corollary}[axiom]{Corollary}
\newtheorem{definition}[axiom]{Definition}
\newtheorem{example}[axiom]{Example}
\newtheorem{lemma}[axiom]{Lemma}
\newtheorem{observation}[axiom]{Observation}
\newtheorem{proposition}[axiom]{Proposition}
\newtheorem{theorem}[axiom]{Theorem}

\newcommand{\QED}{{%
 \parfillskip=0pt
 \widowpenalty=10000
 \displaywidowpenalty=10000
 \finalhyphendemerits=0
 \leavevmode
 \unskip
 \nobreak
 \hfil 
 \penalty50 
 \hskip.2em 
 \null 
 \hfill 
 $\Box$ \par} }

\newenvironment{Poef}[1][]
  {\trivlist \item[\hskip \labelsep{\slshape Proof{#1}:}]\rmfamily }{\endtrivlist}
\newcommand{\proof}[1][]{\begin{Poef}[#1]}
\newcommand{\qed}{\QED\end{Poef}}

\newcommand{\rem}[1]{\marginpar{\addtolength{\baselineskip}{-6pt}{\scriptsize #1}}}

\newcommand{\rz}{{\mathbb{R}}}
\newcommand{\nz}{{\mathbb{N}}}
\newcommand{\zz}{{\mathbb{Z}}}
\newcommand{\eps}{\varepsilon}
\newcommand{\cei}[1]{\lceil #1\rceil}
\newcommand{\flo}[1]{\left\lfloor #1\right\rfloor}
\newcommand{\seq}[1]{\langle #1\rangle}

\newcommand{\mminn}{M_0}
\newcommand{\mmout}{M_{m+1}}
\newcommand{\mmcap}{\mbox{\rm cap}}
\newcommand{\ccc}{{\sc Central Control}}
\newcommand{\bbb}{{\cal B}}
\newcommand{\jjj}{{\cal J}}
\newcommand{\kkk}{{\cal K}}
\newcommand{\mmm}{{\cal M}}
\newcommand{\uuu}{{\cal U}}
\newcommand{\notxi}{\overline{x_i}}

\newcommand{\probSState}{{\sc Safe State Recognition}}
\newcommand{\probRState}{{\sc Reachable State Recognition}}
\newcommand{\probStoS}{{\sc State-to-State Reachability}}
\newcommand{\probDeadlock}{{\sc Reachable Deadlock}}
\newcommand{\probTDM}{{\sc Three-Dimensional Matching}}
\newcommand{\probSAT}{{\sc Three-Satisfiability}}

\title{{\bf Analysis of multi-stage open shop processing systems}
\thanks{This research has been supported
by the Netherlands Organisation for Scientific Research (NWO), grant 639.033.403;
by DIAMANT (an NWO mathematics cluster);
by the Future and Emerging Technologies unit of the European Community (IST priority),
under contract no.\ FP6-021235-2 (project ARRIVAL);
by BSIK grant 03018 (BRICKS: Basic Research in Informatics for Creating the
Know\-ledge Society).}}
\author{\sc Christian E.J. Eggermont
        \thanks{{\tt c.e.j.eggermont@tue.nl}.
        Department of Mathematics and Computer Science,
        TU Eindhoven, P.O.\ Box 513, 5600 MB Eindhoven, Netherlands}
        \and
        \sc Alexander Schrijver
        \thanks{{\tt lex@cwi.nl}. 
        CWI, 1098 XG Amsterdam, and University of Amsterdam, Netherlands}
        \and
        \sc Gerhard J. Woeginger
        \thanks{{\tt gwoegi@win.tue.nl}.
        Department of Mathematics and Computer Science,
        TU Eindhoven, P.O.\ Box 513, 5600 MB Eindhoven, Netherlands}
}
\date{}
\maketitle

\begin{abstract}
We study algorithmic problems in multi-stage open shop processing systems
that are centered around reachability and deadlock detection questions.

We characterize safe and unsafe system states. 
We show that it is easy to recognize system states that can be reached from
the initial state (where the system is empty), but that in general it is hard 
to decide whether one given system state is reachable from another given 
system state.
We show that the problem of identifying reachable deadlock states is hard in 
general open shop systems, but is easy in the special case where no job needs 
processing on more than two machines (by linear programming and matching theory), 
and in the special case where all machines have capacity one (by graph-theoretic
arguments).  

\end{abstract}

\newpage
\section{Introduction}
We consider a \emph{multi-stage open shop} processing system with $n$ jobs
$J_1,\ldots,J_n$ and $m$ machines $M_1,\ldots,M_m$.
Every job $J_j$ ($j=1,\ldots,n$) requests processing on a certain subset $\mmm(J_j)$
of the machines; the ordering in which job $J_j$ passes through the machines in
$\mmm(J_j)$ is irrelevant and can be chosen arbitrarily by the scheduler.
Every machine $M_i$ ($i=1,\ldots,m$) has a corresponding \emph{capacity} $\mmcap(M_i)$,
which means that at any moment in time it can simultaneously hold and process up
to $\mmcap(M_i)$ jobs.
For more information on multi-stage scheduling systems, the reader is referred to the
survey \cite{LLRS}.

In this article, we are mainly interested in the performance of \emph{real-time}
multi-stage systems, where the processing time $p_{j,i}$ of job $J_j$ on machine
$M_i$ is a priori unknown and hard to predict.
The {\ccc} (the scheduling policy) of the system learns the processing time $p_{j,i}$
only when the processing of job $J_j$ on machine $M_i$ is completed.
The various jobs move through the system in an unsynchronized fashion.
Here is the standard behavior of a job in such a system:
\begin{itemize}
\item[1.]
In the beginning the job is asleep and is waiting outside the system.
For technical reasons, we assume that the job occupies an artificial machine
$\mminn$ of unbounded capacity.
\item[2.]
After a finite amount of time the job wakes up, and starts looking for an available
machine $M$ on which it still needs processing.
If the job detects such a machine $M$, it requests permission from the {\ccc} to move
to machine $M$.
If no such machine is available or if the {\ccc} denies permission, the job falls
asleep again (and returns to the beginning of Step~2).
\item[3.]
If the job receives permission to move, it releases its current machine and starts
processing on the new machine $M$.
While the job is being processed and while the job is asleep, it continuously
occupies machine $M$ (and blocks one of the $\mmcap(M)$ available places on $M$).
When the processing of the job on machine $M$ is completed and in case the job still
needs processing on another machine, it returns to Step~2.
\item[4.]
As soon as the processing of the job on all relevant machines is completed, the job
informs the {\ccc} that it is leaving the system.
We assume that the job then moves to an artificial final machine $\mmout$ (with
unbounded capacity), and disappears.
\end{itemize}

The described system behavior typically occurs in robotic cells and flexible
manufacturing systems.
The high level goal of the {\ccc} is to arrive at the situation where all the jobs
have been completed and left the system.
Other goals are of course to reach a high system throughput, and to avoid unnecessary
waiting times of the jobs.
However special care has to be taken to prevent the system from reaching situations
of the following type:
\begin{example}
\label{ex:1}
Consider an open shop system with three machines $M_1,M_2,M_3$ of capacity~1.
There are three jobs that each require processing on all three machines.
Suppose that the {\ccc} behaves as follows:

The first job requests permission to move to machine $M_1$. Permission granted.

The second job requests permission to move to machine $M_2$. Permission granted.

The third job requests permission to move to machine $M_3$. Permission granted.

\noindent
Once the three jobs have completed their processing on theses machines, they keep
blocking their machines and simultaneously keep waiting for the other machines to
become idle.
The processing never terminates.
\end{example}

Example~\ref{ex:1} illustrates a so-called \emph{deadlock}, that is, a situation in
which the system gets stuck and comes to a halt since no further processing is possible:
Every job in the system is waiting for resources that are blocked by other jobs
that are also waiting in the system.
Resolving a deadlock is usually expensive (with respect to time, energy, and resources),
and harmfully diminishes the system performance.
In robotic cells resolving a deadlock typically requires human interaction.
The scientific literature on deadlocks is vast, and touches many different areas like
flexible manufacturing, automated production, operating systems, Petri nets, 
network routing, etc.

The literature distinguishes two basic types of system states (see for instance 
Coffman, Elphick \& Shoshani \cite{CES}, Gold \cite{Gold78}, or Banaszak \&
Krogh \cite{BK90}).
A state is called \emph{safe}, if there is at least one possible way of completing all jobs.
A state is called \emph{unsafe}, if every possible continuation eventually will get stuck
in a deadlock.
An example for a safe state is the initial situation where all jobs are outside the
system (note that the jobs could move sequentially through the system and complete).
Another example for a safe state is the final situation where all jobs have been
completed.
An example for an unsafe state are the deadlock states.

\subsection*{Summary of considered problems and derived results}
\nopagebreak
In this article we study the behavior of safe and unsafe states in open shop scheduling 
systems.
In particular, we investigate the computational complexity of the four algorithmic questions
described in the following paragraphs.
First, if one wants to have a smoothly running system, then it is essential to 
distinguish the safe from the unsafe system states:
\begin{quote}
{\sc Problem:} {\probSState}
\\[1.0ex]
{\sc Instance:}
An open shop scheduling system.
A system state $s$.
\\[1.0ex]
{\sc Question:}
Is state $s$ safe?
\end{quote}
Section~\ref{sec:unsafe} provides a simple characterization of unsafe states, which leads 
to a (straightforward) polynomial time algorithm for telling safe states from unsafe states.
Similar characterizations have already been given a decade ago in the work of Sulistyono 
\& Lawley \cite{SuLa01} and Xing, Lin \& Hu \cite{XiLiHu01}.
Our new argument is extremely short and simple.

One of the most basic problems in analyzing a system consists in characterizing those 
system states that can be reached while the shop is running.
\begin{quote}
{\sc Problem:} {\probRState}
\\[1.0ex]
{\sc Instance:}
An open shop scheduling system.
A system state $s$.
\\[1.0ex]
{\sc Question:}
Can the system reach state $s$ when starting from the initial situation where all 
machines are still empty?
\end{quote}
In Section~\ref{sec:reachability1} we derive a polynomial time algorithm for recognizing 
reachable system states.
The main idea is to reverse the time axis, and to make the system run backward.
Then reachable states in the original system translate into safe states in the reversed 
system, and the results from Section~\ref{sec:unsafe} can be applied.

Hence recognizing states that are reachable from the initial situation is easy.
What about  recognizing states that are reachable from some other given state?
\begin{quote}
{\sc Problem:} {\probStoS}
\\[1.0ex]
{\sc Instance:}
An open shop scheduling system.
Two system states $s$ and $t$.
\\[1.0ex]
{\sc Question:}
Can the system reach state $t$ when starting from state $s$?
\end{quote}
Surprisingly, there is a strong and sudden jump in the computational complexity of the
reachability problem:
Section~\ref{sec:reachability2} provides an NP-hardness proof for problem {\probStoS}.

Another fundamental question is whether an open shop system can ever fall into a deadlock.
In case it cannot, then there are no reachable unsafe states and the {\ccc} may permit
all moves right away and without analyzing them; in other words the system is fool-proof
and will run smoothly without supervision.
\begin{quote}
{\sc Problem:} {\probDeadlock}
\\[1.0ex]
{\sc Instance:}
An open shop scheduling system.
\\[1.0ex]
{\sc Question:}
Can the system ever reach a deadlock state when starting from the initial situation?
\end{quote}
Section~\ref{sec:hard} proves problem {\probDeadlock} to be NP-hard, even for the highly 
restricted special case where the capacity of each machine is at most three and where
each job requires processing on at most four machines.
In Sections \ref{sec:specialcase.1} and \ref{sec:specialcase.2} we exhibit two special 
cases for which this problem is solvable in polynomial time:
The special case where every job needs processing on at most two machines is settled
by a linear programming formulation and techniques from matching theory.
The special case where every machine has capacity one is solved by analyzing cycles 
in certain edge-colored graphs.

\section{Basic definitions}
\nopagebreak
A \emph{state} of an open shop scheduling system is a snapshot describing a
situation that might potentially occur while the system is running.
A state $s$ specifies for every job $J_j$
\begin{itemize}
\item
the machine $M^s(J_j)$ on which this job is currently waiting or currently being
processed,
\item
and the set $\mmm^s(J_j)\subseteq\mmm(J_j)-\{M^s(J_j)\}$ of machines on which the job
still needs future processing.
\end{itemize}
The machines $M^s(J_j)$ implicitly determine
\begin{itemize}
\item
the set $\jjj^s(M_i)\subseteq\{J_1,\ldots,J_n\}$ of jobs currently handled by
machine $M_i$.
\end{itemize}
The \emph{initial state} $0$ is the state where all jobs are still waiting for their
first processing; in other words in the initial state all jobs $J_j$ satisfy
$M^0(J_j)=\mminn$ and $\mmm^0(J_j)=\mmm(J_j)$.
The \emph{final state} $f$ is the state where all jobs have been completed; in other
words in the final state all jobs $J_j$ satisfy $M^f(J_j)=\mmout$ and
$\mmm^f(J_j)=\emptyset$.

A state $t$ is called a \emph{successor} of a state $s$, if it results from $s$ by
moving a single job $J_j$ from its current machine $M^s(J_j)$ to some new machine in
set $\mmm^s(J_j)$, or by moving a job $J_j$ with $\mmm^s(J_j)=\emptyset$ from its
current machine to $\mmout$.
In this case we will also say that the system \emph{moves} from $s$ to $t$.
This successor relation is denoted $s\to t$.
A state $t$ is said to be \emph{reachable} from state $s$, if there exists a finite
sequence $s=s_0,s_1,\ldots,s_k=t$ of states (with $k\ge0$) such that $s_{i-1}\to s_i$
holds for $i=1,\ldots,k$.
A state $s$ is called \emph{reachable}, if it is reachable from the initial state 0.

\begin{proposition}
\label{pr:finite}
Any reachable state $s$ can be reached from the initial state through a sequence
of at most $n+\sum_{i=1}^n|\mmm(J_j)|$ moves.
\end{proposition}

A state is called \emph{safe}, if the final state $f$ is reachable from it;
otherwise the state is called \emph{unsafe}.
A state is a \emph{deadlock}, if it has no successor states and if it is not the
final state $f$.

\section{Analysis of unsafe states}
\label{sec:unsafe}
Unsafe states in open shop systems are fairly well-understood, and the literature 
contains several characterizations for them; see for instance 
Sulistyono \& Lawley \cite{SuLa01}, Xing, Lin \& Hu \cite{XiLiHu01}, and
Lawley \cite{Lawley99}.
In this section we provide yet another analysis of unsafe states, which is shorter 
and (as we think) simpler than the previously published arguments.

A machine $M$ is called \emph{full} in state $s$, if it is handling exactly $\mmcap(M)$ jobs.
A non-empty subset $\bbb$ of the machines is called \emph{blocking} for state $s$,
\begin{itemize}
\item if every machine in $\bbb$ is full, and
\item if every job $J_j$ that occupies some machine in $\bbb$ satisfies
$\emptyset\ne\mmm^s(J_j)\subseteq\bbb$.
\end{itemize}

Here is a simple procedure that determines whether a given machine $M_i$ is part of
a blocking set in state $s$:
Let $\bbb_0=\{M_i\}$.
For $k\ge1$ let $\jjj_k$ be the union of all job sets $\jjj^s(M)$ with $M\in\bbb_{k-1}$,
and let $\bbb_k$ be the union of all machine sets $\mmm^s(J)$ with $J\in\jjj_k$.
Clearly $\bbb_0\subseteq\bbb_1\subseteq\cdots\subseteq\bbb_{m-1}=\bbb_m$.
Furthermore machine $M_i$ belongs to a blocking set, if and only if $\bbb_m$ is a
blocking set, if and only if all machines in $\bbb_m$ are full.
In case $\bbb_m$ is a blocking set, we denote it by $\bbb^s_{\min}(M_i)$ and call it
the \emph{canonical} blocking set for machine $M_i$ in state $s$.
The canonical blocking set is the smallest blocking set containing $M_i$:
\begin{lemma}
\label{le:blocking.1}
If machine $M_i$ belongs to a blocking set $\bbb$ in state $s$, then
$\bbb^s_{\min}(M_i)\subseteq\bbb$.
\end{lemma}

The machines in a blocking set $\bbb$ all operate at full capacity on jobs that
in the future only want to move to other machines in $\bbb$.
Since these jobs are permanently blocked from moving, the state $s$ must eventually
lead to a deadlock and hence is unsafe.
The following theorem shows that actually \emph{every} deadlock is caused by
such blocking sets.

\begin{theorem}
\label{th:blocking}
A state $s$ is unsafe if and only if it has a blocking set of machines.
\end{theorem}
\proof
The if-statement is obvious.
For the only-if-statement, we classify the unsafe states with respect to their 
distances to deadlock states.
The set $\uuu_0$ contains the deadlock states.
For $d\ge1$, set $\uuu_d$ contains all states whose successor states are all
contained in $\uuu_{d-1}$.
Note that $\uuu_{d-1}\subseteq \uuu_d$, and note that every unsafe state occurs in
some $\uuu_d$.
We prove by induction on $d$ that every state in $\uuu_d$ has a blocking set of machines.
For $d=0$ this is trivial.

In the inductive step, assume for the sake of contradiction that some state $s\in\uuu_d$
is unsafe but does not contain any blocking set.
Since every move from $s$ leads to a state in $\uuu_{d-1}$, all successor states of $s$
must contain blocking sets.
Whenever in state $s$ some job $J$ moves to some (non-full) machine $M$, this machine $M$
must become full and must then be part of any blocking set.
Among all possible moves, consider a move that yields a state $t$ with a newly full
machine $M$ for which the canonical blocking set $\bbb^t_{\min}(M)$ is of the smallest
possible cardinality.

Note that in state $t$ there exist a machine $M'\in\bbb^t_{\min}(M)$ and a job
$J'\in\jjj^t(M')$ with $M\in\mmm^t(J')$; otherwise $\bbb^t_{\min}(M)-\{M\}$ would be
a blocking set for state $s$.
Now consider the successor state $u$ of $s$ that results by moving job $J'$ from machine
$M$ to $M'$.
Since $\mmm^u(J')\subseteq\bbb^t_{\min}(M)$, a simple inductive argument shows that
$\bbb^u_{\min}(M)\subseteq\bbb^t_{\min}(M)$.
Since job $J'$ has just jumped away from $M'$, this machine cannot be full in state $u$,
and hence $M'\in\bbb^t_{\min}(M)-\bbb^u_{\min}(M)$.
Consequently the canonical blocking set $\bbb^u_{\min}(M)$ has smaller cardinality
than $\bbb^t_{\min}(M)$.
This contradiction completes the proof.
\qed

\begin{lemma}
\label{le:blocking.algo}
For a given state $s$, it can be decided in polynomial time whether $s$ has a
blocking set of machines.
Consequently, problem {\probSState} can be decided in polynomial time.
\end{lemma}
\proof
Create an auxiliary digraph that corresponds to state $s$:
the vertices are the machines $M_1,\ldots,M_m$.
Whenever some job $J_j$ occupies a machine $M_i$, the digraph contains an arc from 
$M_i$ to every machine in $\mmm^s(J_j)$.
Obviously state $s$ has a blocking set of machines if and only if the auxiliary
digraph contains a strongly connected component with the following two properties:
(i) All vertices in the component are full.
(ii) There are no arcs leaving the component.
Since the strongly connected components of a digraph can easily be determined and
analyzed in linear time (see for instance \cite{CLRS}), the desired statement follows.
\qed

\section{Analysis of reachable states}
\label{sec:reachability1}
In this section we discuss the behavior of reachable system states.
We say that a state $t$ is \emph{subset-reachable} from state $s$, if every job $J_j$ 
satisfies one of the following three conditions:
\begin{itemize}
\item $M^t(J_j)=M^s(J_j)$ and $\mmm^t(J_j)=\mmm^s(J_j)$, or
\item $M^t(J_j)\in\mmm^s(J_j)$ and $\mmm^t(J_j)\subseteq\mmm^s(J_j)-\{M^t(J_j)\}$, or
\item $M^t(J_j)=\mmout$ and $\mmm^t(J_j)=\emptyset$.
\end{itemize}
Clearly whenever a state $t$ is reachable from some state $s$, then $t$ is also
subset-reachable from $s$.
The following example demonstrates that the reverse implication is not necessarily true.
This example also indicates that the algorithmic problem {\probRState} (as formulated
in the introduction) is not completely straightforward.

\begin{example}
\label{ex:unreachable}
Consider an open shop system with two machines $M_1,M_2$ of capacity~1 and
two jobs $J_1,J_2$ with $\mmm(J_1)=\mmm(J_2)=\{M_1,M_2\}$.
Consider the state $s$ where $J_1$ is being processed on $M_1$ and $J_2$ is
being processed on $M_2$, and where $\mmm^s(J_1)=\mmm^s(J_2)=\emptyset$.
It can be seen that $s$ is subset-reachable from the initial state 0,
whereas $s$ is not reachable from 0.
\end{example}

Our next goal is to derive a polynomial time algorithm for recognizing reachable
system states.
Consider an open shop scheduling system and a fixed system state $s$.
Without loss of generality we assume that $s$ is subset-reachable from the initial state.
We define a new (artificial) state $t$ where $M^t(J_j):=M^{s}(J_j)$ and
$\mmm^t(J_j):=\mmm(J_j)-\mmm^s(J_j)-\{M^s(J_j)\}$ for all jobs $J_j$.
Note that in both states $s$ and $t$ every job is sitting on the very same machine, 
but the work that has already been performed in state $s$ is exactly the work that 
still needs to be done in state $t$.

\begin{lemma}
\label{le:reach.reverse}
State $s$ is reachable if and only if state $t$ is safe.
\end{lemma}
\proof
First assume that $s$ is reachable, and let $0=s_0\to s_1\to\cdots\to s_k=s$ denote
a corres\-ponding witness sequence of moves.
Define a new sequence $t=t_k\to t_{k-1}\to\cdots\to t_0=f$ of moves:
Whenever the move $s_{\ell}\to s_{\ell+1}$ ($0\le\ell\le k-1$) results from moving
job $J_j$ from machine $M_a$ to machine $M_b$, then the move $t_{\ell+1}\to t_{\ell}$
results from moving job $J_j$ from machine $M_b$ to machine $M_a$.
(Note that the artificial machines $\mminn$ and $\mmout$ switch their roles.)
Hence $t$ is safe.
A symmetric argument shows that if $t$ is safe then $s$ is reachable.
\qed

Hence deciding reachability is algorithmically equivalent to deciding safeness.
Together with Lemma~\ref{le:blocking.algo} this yields the following theorem.
\begin{theorem}
\label{th:reachability}
{\probRState} can be decided in polynomial time.  \QED
\end{theorem}

The following lemma states a simple sufficient condition that makes a state reachable.
\begin{lemma}
\label{le:reach.sufficient}
Let $s$ be a state, and let $\kkk$ be a subset of machines such that every job that
still needs further processing in $s$ satisfies $M^s(J_j)\in\kkk$ and
\[ \mmm^s(J_j)\cup\{M^s(J_j)\} ~=~ \kkk\cap\mmm(J_j).\]
Then $s$ is a reachable system state.
\end{lemma}
\proof
By renaming the jobs we assume that the jobs $J_j$ with $1\le j\le k$ have
$M^s(J_j)=\mmout$ and the jobs $J_j$ with $k+1\le j\le n$ have $M^s(J_j)\in\kkk$.
We handle the jobs one by one in their natural order:
every job moves through all machines in $\mmm(J_j)-\mmm^s(J_j)$, and ends up on 
machine $M^s(J_j)$.
Then the next job is handled.
\qed

\section{Analysis of state-to-state reachability}
\label{sec:reachability2}
We establish NP-hardness of {\probStoS} by means of a reduction from the following 
satisfiability problem; see Garey \& Johnson \cite{GaJo79}.
\begin{quote}
{\sc Problem:} {\probSAT}
\\[1.0ex]
{\sc Input:}
A set $X=\{x_1,\ldots,x_n\}$ of $n$ logical variables;
a set $C=\{c_1,\ldots,c_m\}$ of $m$ clauses over $X$ that each contain three literals.
\\[1.0ex]
{\sc Question:}
Is there a truth assignment for $X$ that satisfies all clauses in $C$?
\end{quote}
We start from an instance of {\probSAT}, and construct a corres\-ponding instance of 
{\probStoS} for it.
Throughout we will use $\ell_i$ to denote the unnegated literal $x_i$ or the negated 
literal $\notxi$ for some fixed variable $x_i\in X$, and we will use $\ell$ to denote 
a generic literal over $X$.
Altogether there are $5n+m$ machines:
\begin{itemize}
\itemsep=0.0ex
\item
For every literal $\ell_i$, there are three corresponding machines $S(\ell_i)$, 
$T(\ell_i)$, and $U(\ell_i)$.
Machine $U(\ell_i)$ has capacity~2, whereas machines $S(\ell_i)$ and $T(\ell_i)$ have 
capacity~1.
For every variable $x_i\in X$ the two machines $U(x_i)$ and $U(\notxi)$ coincide,
and the corresponding machine will sometimes simply be called $U(i)$.
\item
For every clause $c_j\in C$, there is a corresponding machine $V(c_j)$ with capacity 3.
\end{itemize}
Furthermore the scheduling instance contains $4n$ jobs that correspond to literals 
and $6m$ jobs that correspond to clauses.
For every literal $\ell_i$ there are two corresponding jobs:
\begin{itemize}
\itemsep=0.0ex
\item
Job $J(\ell_i)$ is sitting on machine $S(\ell_i)$ in state $s$. 
In state $t$ it has moved to machine $U(\ell_i)$ without visiting other machines 
inbetween.
\item
Job $J'(\ell_i)$ is still waiting outside the system in state $s$, and has already 
left the system in state $t$.
Inbetween the job visits machines $S(\ell_i)$, $T(\ell_i)$, $U(\ell_i)$ in arbitrary
order.
\end{itemize}
Consider a clause $c_j$ that consists of three literals $\ell_a,\ell_b,\ell_c$.
Then the following six jobs correspond to clause $c_j$:
\begin{itemize}
\itemsep=0.0ex
\item
For $\ell\in\{\ell_a,\ell_b,\ell_c\}$ there is a job $K(c_j,\ell)$ that in state $s$
sits on machine $V(c_j)$, then moves through machines $S(\ell)$ and $T(\ell)$
in arbitrary order, and finally has left the system in state $t$.  
Note that in state $s$ these three jobs block machine $V(c_j)$ to full capacity.
\item
For $\ell\in\{\ell_a,\ell_b,\ell_c\}$ there is another job $K'(c_j,\ell)$ that waits
outside the system in state $s$, then moves through machines $U(\ell)$ and $V(c_j)$
in arbitrary order, and finally has left the system in state $t$.
\end{itemize}
In Sections \ref{subsec:dag.1} and \ref{subsec:dag.2} we will show that in the
constructed scheduling instance state $t$ is reachable from state $s$
if and only if the {\probSAT} instance has a satisfying truth assignment.
This then implies the following theorem.
\begin{theorem}
\label{th:state-to-state}
{\probStoS} is NP-complete.
\end{theorem}

\subsection{Proof of the if-statement}
\label{subsec:dag.1}
We assume that the {\probSAT} instance has a satisfying truth assignment.
We describe a sequence of moves that brings the scheduling system from the starting
state $s$ into the goal state $t$.

In a first phase, for every true variable $x_i$ the job $J(x_i)$ moves from machine 
$S(x_i)$ to machine $U(i)$.
Then job $J'(x_i)$ enters the system by moving to $U(i)$, then moves to $T(x_i)$, 
then to $S(x_i)$, and finally leaves the system.
Next job $J'(\notxi)$ enters the system, moves to $U(i)$, and finally sits and waits 
on $T(\notxi)$.
Symmetric moves (with the roles of $x_i$ and $\notxi$ interchanged) are performed 
for every false variable $x_i$.
At the end of this phase, for every true literal $\ell_i$ the two machines $S(\ell_i)$ 
and $T(\ell_i)$ are empty, and there is an empty spot on machine $U(\ell_i)$.

In the second phase, we consider clauses $c_j$ that consist of three literals 
$\ell_a,\ell_b,\ell_c$.
We pick one true literal $\ell_i$ from $c_j$, and we let the corresponding job 
$K(c_j,\ell_i)$ jump away from machine $V(c_j)$ to machine $S(\ell_i)$, then to 
$T(\ell_i)$, and finally make it leave the system.
This yields a free spot on machine $V(c_j)$.
For every $\ell\in\{\ell_a,\ell_b,\ell_c\}$ we let job $K'(c_j,\ell)$ enter the
system, move through machines $U(\ell)$ and $V(c_j)$, and then leave the system.
At the end of this phase, four out of the six jobs corresponding to every clause have
reached their final destination in state $t$.

In the third phase, for every true variable $x_i$ the job $J(\notxi)$ moves from
machine $S(\notxi)$ to machine $U(i)$.
Job $J'(\notxi)$ moves from $T(\notxi)$ to $S(\notxi)$, and then leaves the system.
Symmetric moves (with the roles of $x_i$ and $\notxi$ interchanged) are performed
for every false variable $x_i$.
At the end of this phase, all jobs $J(\ell_i)$ and $J'(\ell_i)$ have reached their
final destination in state $t$.
All machines $S(\ell_i)$ and $T(\ell_i)$ are empty.

In the fourth phase, we again consider clauses $c_j$ that consist of three literals 
$\ell_a,\ell_b,\ell_c$.
For the two literals $\ell$ in $c_j$ that did not get picked in the second phase,
we move the corresponding job $K(c_j,\ell)$ from machine $V(c_j)$ to machine $S(\ell)$, 
then to machine $T(\ell)$, and finally make it leave the system.
At the end of this phase all jobs have reached their final destination, and the
system has reached the desired goal state $t$.

\subsection{Proof of the only-if-statement}
\label{subsec:dag.2}
We assume that there is a sequence of moves that brings the scheduling system 
from state $s$ into state $t$.
We will deduce from this a satisfying truth assignment for the {\probSAT} instance.

We say that variable $x_i$ is \emph{activated} as soon as one of the corresponding
jobs $J(x_i)$ and $J(\notxi)$ moves to machine $U(i)$. 
We say that $x_i$ is \emph{deactivated} at the moment $\mu_i$ in time where also 
the other job $J(x_i)$ or $J(\notxi)$ moves to machine $U(i)$.
If job $J(x_i)$ moves first and activates $x_i$, we set variable $x_i$ to true; 
if $J(\notxi)$ moves first and activates $x_i$, we set variable $x_i$ to false.
We will show that the resulting truth setting satisfies all clauses.

\begin{lemma}
\label{le:dag.x1}
If $\ell_i$ is a false literal, then job $K(c_j,\ell_i)$ can visit machine 
$T(\ell_i)$ only after the deactivation time $\mu_i$ of variable $x_i$.
\end{lemma}
\proof
Till the crucial moment $\mu_i$ where variable $x_i$ is deactivated, job 
$J(\ell_i)$ is permanently blocking machine $S(\ell_i)$.
{From} time $\mu_i$ onwards, jobs $J(x_i)$ and $J(\notxi)$ together are permanently 
blocking the machine $U(i)$ with capacity~2.

Suppose for the sake of contradiction that some job $K(c_j,\ell_i)$ moves to machine 
$T(\ell_i)$ before moment $\mu_i$. 
Then at time $\mu_i$, it is waiting for its final processing on machine $S(\ell_i)$
and blocking machine $T(\ell_i)$.
We claim that under these circumstances job $J'(\ell_i)$ is causing trouble:
In case $J'(\ell_i)$ has not yet entered the system at time $\mu_i$, it can never be 
processed on machine $U(i)$ which is permanently blocked from time $\mu_i$ onwards.
In case $J'(\ell_i)$ has already entered the system at time $\mu_i$, then at time 
$\mu_i$ it must be sitting on machine $U(i)$ and thereby prevents job $J(\ell_i)$ 
from moving there.
In either case we reach a contradiction.
\qed

Now let us consider some arbitrary clause $c_j$ that consists of three literals 
$\ell_a,\ell_b,\ell_c$, let $x_a,x_b,x_c$ be the three underlying variables in $X$, 
and assume without loss of generality that the corresponding moments of deactivation 
satisfy $\mu_a<\mu_b<\mu_c$.
\begin{lemma}
\label{le:dag.x2}
At time $\mu_a$ job $K'(c_j,\ell_a)$ must either be sitting on machine $V(c_j)$, 
or must have left the system.
\end{lemma}
\proof
If at time $\mu_a$ job $K'(c_j,\ell_a)$ is still waiting outside the system,
then it will never be processed on machine $U(a)$, which is permanently blocked 
by jobs $J(x_a)$ and $J(\overline{x_a})$.
Hence there is no way of reaching state $t$, which is a contradiction.
If at time $\mu_a$ job $K'(c_j,\ell_a)$ is sitting on machine $U(a)$, it thereby
prevents variable $x_a$ from being deactivated.
That's another contradiction.
\qed

Hence at time $\mu_a$ job $K'(c_j,\ell_a)$ must already have visited machine $V(c_j)$.
Since in the starting state $s$ the three jobs $K(c_j,\ell_a)$, $K(c_j,\ell_b)$, 
$K(c_j,\ell_c)$ are blocking $V(c_j)$, one of them must have made space and must have 
moved away before time $\mu_a$; let this job be $K(c_j,\ell_i)$ where $i\in\{a,b,c\}$.
We distinguish two cases.
First, assume that $K(c_j,\ell_i)$ has moved to machine $S(\ell_i)$. 
Since variable $x_i$ is still active, literal $\ell_i$ must be true in this case.
Secondly, assume that $K(c_j,\ell_i)$ has moved to machine $T(\ell_i)$.
Since variable $x_i$ is still active, Lemma~\ref{le:dag.x1} yields that literal 
$\ell_i$ is true.
In either case, clause $c_j$ contains the true literal $\ell_i$.

We conclude that every clause contains some true literal, and that the defined truth
setting satisfies all clauses.
This completes the proof of Theorem~\ref{th:state-to-state}.

\section{Analysis of reachable deadlocks}
\label{sec:hard}
In this section we show that {\probDeadlock} is an NP-hard problem.
Our reduction is from the following variant of the {\probTDM} problem;
see Garey \& Johnson \cite[p.221]{GaJo79}.
\begin{quote}
{\sc Problem:} {\probTDM}
\\[1.0ex]
{\sc Instance:}
An integer $n$.
Three pairwise disjoint sets $A=\{a_1,\ldots,a_n\}$, $B=\{b_1,\ldots,b_n\}$, and 
$C=\{c_1,\ldots,c_n\}$.
A set $T\subseteq A\times B\times C$ of triples, such that every element occurs
in at most three triples in $T$.
\\[1.0ex]
{\sc Question:}
Does there exist a subset $T'\subseteq T$ of $n$ triples, such that every element in
$A\cup B\cup C$ occurs in exactly one triple in $T'$?
\end{quote}
We start from an arbitrary instance of {\probTDM}, and construct the following
corresponding instance of {\probDeadlock} for it.
There are two types of machines.
Note that every machine has capacity at most three.
\begin{itemize}
\itemsep=0.0ex
\item
There are $n+2$ so-called \emph{structure machines} $S_0,\ldots,S_{n+1}$, each of capacity~1.
\item
For every triple $t\in T$, there is a corresponding \emph{triple machine} $T_t$ with capacity 3.
\end{itemize}
Furthermore there are $4n+2$ jobs.
\begin{itemize}
\itemsep=0.0ex
\item
For every element $a_i\in A$ there are two corresponding \emph{A-element jobs}
$J^+(a_i)$ and $J^-(a_i)$.
Job $J^+(a_i)$ requires processing on structure machine $S_i$, and on every triple
machine $T_t$ with $a_i\in t$.
Job $J^-(a_i)$ requires processing on structure machine $S_{i-1}$, and on every triple
machine $T_t$ with $a_i\in t$.
\item
For every element $b_i\in B$ there is a corresponding \emph{B-element job} $J(b_i)$
that requires processing on structure machine $S_{n+1}$, and on every triple machine
$T_t$ with $b_i\in t$.
\item
For every element $c_i\in C$ there is a corresponding \emph{C-element job} $J(c_i)$
that requires processing on structure machine $S_{n+1}$, and on every triple machine
$T_t$ with $c_i\in t$.
\item
Finally there is a dummy job $D_0$ that needs processing on $S_0$ and $S_{n+1}$,
and another dummy job $D_{n+1}$ that needs processing on $S_n$ and $S_{n+1}$.
\end{itemize}
Since every element of $A\cup B\cup C$ occurs in at most three triples, we note that
each job requires processing on at most four machines.
For the ease of later reference, we also list for every machine the jobs that need
processing on that machine.
\begin{itemize}
\itemsep=0.0ex
\item
A triple machine $T_t$ with $t=(a_i,b_j,c_k)$ handles the four jobs $J^+(a_i)$,
$J^-(a_i)$, $J(b_j)$, and $J(c_k)$.
\item
Structure machine $S_i$ with $1\le i\le n-1$ handles the jobs $J^+(a_i)$ and
$J^-(a_{i+1})$.
\\
Structure machine $S_0$ handles the two jobs $J^-(a_1)$ and $D_0$. 
\\
Structure machine $S_n$ handles the two jobs $J^+(a_n)$ and $D_{n+1}$.
\\
Structure machine $S_{n+1}$ handles $2n+2$ jobs:
$D_0$, $D_{n+1}$, all B-element jobs, and all C-element jobs.
\end{itemize}
The following theorem contains the main result of this section.
\begin{theorem}
\label{th:reachable-deadlock}
{\probDeadlock} is NP-complete, even if the capacity of each machine is at most three,
and  if each job requires processing on at most four machines.  
\end{theorem}

Indeed, Proposition~\ref{pr:finite} yields an NP-certificate for problem {\probDeadlock}.
The hardness argument proves that the constructed scheduling instance has a reachable 
deadlock if and only if the {\probTDM} instance has answer YES.
The only-if-statement will be proved in Section~\ref{subsec:hard.1}, and the
if-statement will be proved in Section~\ref{subsec:hard.2}.

\subsection{Proof of the only-if-statement}
\label{subsec:hard.1}
We assume that the scheduling instance has a reachable deadlock state $s$, and we
will show that then the {\probTDM} instance has answer YES.

Let $\bbb^*$ be a blocking set of minimum cardinality in $s$, and let $\jjj^*$ denote
the jobs that are currently being processed on machines in $\bbb^*$.
For every triple machine $T_t$ in $\bbb^*$, the job set $\jjj^*$ contains all four
element jobs that need processing on $T_t$. (First: Machine $T_t$ must be full and hence
must process three jobs. Second: If no other job in $\jjj^*$ needs processing on $T_t$,
then $\bbb^*-\{T_t\}$ would yield a smaller blocking set.)
Similarly, for every structure machine $S_i\in\bbb^*$ with $0\le i\le n$, the job set
$\jjj^*$ contains both jobs that need processing on $S_i$.

\begin{lemma}
\label{le:hard.x1}
The blocking set $\bbb^*$ contains at least one of the structure machines $S_i$
with $0\le i\le n$.
\end{lemma}
\proof
Suppose otherwise.
Then $\bbb^*$ consists solely of triple machines and perhaps of machine $S_{n+1}$.

We first claim that every triple machine in $\bbb^*$ processes exactly one
A-element job, one B-element job, and one C-element job.
Indeed, there is an A-element job $J\in\jjj^*$ that corresponds to some element 
$a_i\in A$ and that is processed on some triple machine $T_t\in\bbb^*$.
The machine set $\mmm^s(J)$ of this job contains another triple machine $T_u\in\bbb^*$.
Then $a_i\in t$ and $a_i\in u$, and both machines $T_t$ and $T_u$ must be processing one 
A-element job (that corresponds to element $a_i$), one B-element job, and one C-element job.
This established the claim.

Next fix a B-element job $J(b_i)\in\jjj^*$ that is processed on some machine $T_t$ in $\bbb^*$.
The machine set $\mmm^s(J(b_i))$ contains yet another machine from $\bbb^*$.
This cannot be a triple machine $T_v\in\bbb^*$. (Every such machine $T_v$ is processing
another B-element jobs $J(b_j)$ with $j\ne i$, which implies $b_i\notin u$).
Hence $J(b_i)$ needs future processing on the structure machine $S_{n+1}$, and 
$S_{n+1}\in\bbb^*$.
Then $S_{n+1}$ must be blocked by some job that needs future processing on some 
other machine in $\bbb^*$.
But neither $D_0$, nor $D_{n+1}$, nor any B-element or C-element job can do that.
\qed

\begin{lemma}
\label{le:hard.x2}
(i) Let $S_i\in\bbb^*$ with $1\le i\le n$, and let job $J^+(a_i)$ be running on $S_i$.
Then there exists exactly one triple machine $T_t\in\bbb^*$ with $a_i\in t$, 
and this machine is processing job $J^-(a_i)$.
Furthermore $S_{i-1}\in\bbb^*$.

(ii) Let $S_{i-1}\in\bbb^*$ with $1\le i\le n$, and let job $J^-(a_i)$ be running on $S_{i-1}$.
Then there exists exactly one triple machine $T_t\in\bbb^*$ with $a_i\in t$, 
and this machine is processing job $J^+(a_i)$.
Furthermore $S_i\in\bbb^*$.
\end{lemma}
\proof
As the statements (i) and (ii) are symmetric, we only discuss (i).
Consider job $J^+(a_i)$ on machine $S_i$.
Since $\emptyset\ne\mmm^s(J^+(a_i))\subseteq\bbb^*$, we conclude that $J^+(a_i)$
still needs to be processed on a triple machine $T_t\in\bbb^*$, say with $t=(a_i,b_j,c_k)$.
Since $T_t$ is full, it must be processing the three jobs $J^-(a_i)$, $J(b_j)$,
and $J(c_k)$.
Then none of the remaining triple machines $T_u$ with $a_i\in u$ can be full, 
and hence none of them can be in $\bbb^*$.

The job $J^-(a_i)\in\jjj^*$ is running on $T_t\in\bbb^*$ and still needs future
processing on another machine in $\bbb^*$.
The only remaining candidate for this machine is $S_{i-1}$.
\qed

\begin{lemma}
\label{le:hard.x3}
The blocking set $\bbb^*$ either contains machine $S_0$ which is busy with 
$D_0\in\jjj^*$, or machine $S_n$ which is busy with $D_{n+1}\in\jjj^*$.
In either case, the blocking set $\bbb^*$ contains the structure machine $S_{n+1}$.
\end{lemma}
\proof
Lemma~\ref{le:hard.x1} yields that $S_r\in\bbb^*$ for some $r$ with $0\le r\le n$.
First assume $1\le r\le n$ and that $S_r$ is busy with $J^+(a_r)$.
Then an inductive argument based on Lemma~\ref{le:hard.x2}.(i) yields $S_i\in\bbb^*$
for $0\le i\le r$.
Moreover for $1\le i\le r$ machine $S_i$ is busy with $J^+(a_i)$, and finally
$S_0\in\bbb^*$ must be busy with $D_0$.
Next assume $0\le r\le n-1$ and that $S_r$ is busy with $J^-(a_{r+1})$.
Then a symmetric argument based on Lemma~\ref{le:hard.x2}.(ii) yields that machine
$S_n\in\bbb^*$ is busy with $D_{n+1}$.
This establishes the first part of the lemma.

If $S_0\in\bbb^*$ is busy with $D_0$, then $D_0\in\jjj^*$ requires future processing 
on another machine in $\bbb^*$, which must be $S_{n+1}$. 
If $S_n\in\bbb^*$ is busy with $D_{n+1}$, then $D_{n+1}\in\jjj^*$ requires future 
processing on another machine in $\bbb^*$, which must be $S_{n+1}$. 
In either case this yields the second part of the lemma.
\qed

{From} now on we will assume that machine $S_0\in\bbb^*$ is busy with $D_0$.
(The case where $S_n\in\bbb^*$ is busy with $D_{n+1}$ can be settled in a symmetric way.)
We distinguish two cases on the job running on $S_{n+1}$.

(Case 1) 
Assume that $S_{n+1}$ is busy with $D_{n+1}\in\jjj^*$.
Then $D_{n+1}$ is waiting for another machine in $\bbb^*$, which must be
machine $S_n$ that is busy with $J^+(a_n)$.
We claim for $1\le i\le n$ that $J^+(a_i)$ is processed on machine $S_i\in\bbb^*$,
and that job $J^-(a_i)$ is processed on a triple machine $T_t\in\bbb^*$ with $a_i\in t$.
The claim is proved by a simple inductive argument based on Lemma~\ref{le:hard.x2}.(i) 
starting with $i=n$ and going down to $i=1$.
Then every triple machine in $\bbb^*$ processes one of the jobs $J^-(a_1),\ldots,J^-(a_n)$,
one B-element job, and one C-element job.
These $n$ triple machines induce a solution for the {\probTDM} instance.

(Case 2)
Assume that $S_{n+1}$ is busy with a B-element job or a C-element job; without loss of 
generality we assume that it is busy with a B-element job $J(b_r)$.
Then $J(b_r)$ is waiting for another machine in $\bbb^*$, which must be a full
triple machine $M_t$ with $t=(a_j,b_r,c_k)$.
Machine $M_t$ is busy with the three jobs $J^-(a_j)$, $J^+(a_j)$, and $J(c_k)$.
\begin{itemize}
\itemsep=0.0ex
\item[(i)]
Job $J^-(a_j)$ is waiting for a full machine in $\bbb^*$. 
If $j\ge2$, then this must be the structure machine $S_{j-1}$ which is 
processing job $J^+(a_{j-1})$ (and if $j=1$, then it is the structure machine $S_0$
which is processing job $D_0$).
An inductive argument based on Lemma~\ref{le:hard.x2}.(i) yields that for $1\le i\le j-1$
job $J^+(a_i)$ is processed on machine $S_i\in\bbb^*$, and job $J^-(a_i)$ is processed 
on a triple machine $T_t\in\bbb^*$ with $a_i\in t$.
\item[(ii)]
Also job $J^+(a_j)$ is waiting for a full machine in $\bbb^*$.
If $j\le n-1$, then this must be the structure machine $S_j$ which is processing 
job $J^-(a_{j+1})$ (and if $j=n$, then it is the structure machine $S_n$ which
is processing job $D_{n+1}$).
An inductive argument based on Lemma~\ref{le:hard.x2}.(ii) yields that for 
$j+1\le i\le n$ job $J^-(a_i)$ is processed on machine $S_{i-1}\in\bbb^*$, and 
job  $J^+(a_i)$ is processed on a triple machine $T_t\in\bbb^*$ with $a_i\in t$.  
\end{itemize}
Now the $j-1$ triple machines in (i), the $n-j$ triple machines in (ii), and the 
triple machine $M_t$ with $t=(a_j,b_r,c_k)$ together induce a solution for the 
{\probTDM} instance.
This completes the analysis of Case~2, and it also completes the proof of 
the only-if-statement.

\subsection{Proof of the if-statement}
\label{subsec:hard.2}
We assume that the {\probTDM} instance has a solution $T'\subseteq T$, and from
this we will derive a reachable deadlock state for the scheduling instance.

Consider the subset $\kkk=\{T_t:t\in T'\}\cup\{S_i:0\le i\le n+1\}$ of machines.
We construct a state $t$ where every job $J$ has already entered the system, has already
been processed on all machines in $\mmm(J)-\kkk$, and is currently being processed on
its first machine from $\mmm(J)\cap\kkk$.
Hence the assignment of jobs to machines determines the entire state $t$.
We assign job $D_0$ to machine $S_0$, and job $D_{n+1}$ to structure machine $S_{n+1}$.
For every triple $t=(a_i,b_j,c_k)\in T'$, we assign the three jobs $J^-(a_i)$,
$J(b_j)$, $J(c_k)$ to triple machine $T_t$, and we assign job $J^+(a_i)$ to
triple machine $S_i$.

The resulting state $t$ has $\kkk$ as blocking set and is in deadlock.
Furthermore Lemma~\ref{le:reach.sufficient} shows that $t$ is a reachable state.
All in all, this yields a reachable deadlock state $t$.

\section{Reachable deadlocks if jobs require two machines}
\label{sec:specialcase.1}
Throughout this section we only consider open shop systems where $|\mmm(J)|=2$
holds for all jobs $J$.
We introduce for every job $J$ and for every machine $M\in\mmm(J)$ a corresponding
real variable $x(J,M)$, and for every machine $M$ a corresponding real variable $y(M)$.
Our analysis is centered around the following linear program (LP):
\[
\begin{array}{lll}
\min &\sum_{M}~ \max\{y(M),~ \mmcap(M) \}
\\[2.5ex] \mbox{s.t.}
     &\sum_{J:M\in\mmm(J)}~ x(J,M) ~=~ y(M) &\mbox{for all machines $M$}
\\[1.5ex]
     &\sum_{M\in\mmm(J)}~ x(J,M) ~=~ 1      &\mbox{for all jobs $J$}
\\[1.5ex]
     &x(J,M) ~\ge~ 0                        &\mbox{for all $J$ and $M\in\mmm(J)$}
\end{array}
\]
Although this linear program is totally unimodular, we will mainly deal with
its fractional solutions.

\begin{lemma}
\label{le:lex.0}
One can compute in polynomial time an optimal solution for the linear program (LP) 
that additionally satisfies the following property (*) for every job $J$ with 
$\mmm(J)=\{M_a,M_b\}$:
If $y(M_a)\ge\mmcap(M_a)$ and $x(J,M_a)>0$, then $y(M_b)\ge\mmcap(M_b)$.
\end{lemma}
\proof
We determine in polynomial time an optimal solution of (LP).
Then we perform a polynomial number of post-processing steps on this optimal solution, 
as long as there exists a job violating property (*).
In this case $y(M_a)\ge\mmcap(M_a)$, $x(J,M_a)>0$, and $y(M_b)<\mmcap(M_b)$.

The post-processing step decreases the values $x(J,M_a)$ and $y(M_a)$ by some 
$\eps>0$, and simultaneously increases $x(J,M_b)$ and $y(M_b)$ by the same $\eps$.
By picking $\eps$ smaller than the minimum of $\mmcap(M_b)-y(M_b)$ and $x(J,M_a)$ 
this will yield another feasible solution for (LP).
What happens to the objective value?
If $y(M_a)>\mmcap(M_a)$ at the beginning of the step, then the step would decrease
the objective value, which contradicts optimality.
If $y(M_a)=\mmcap(M_a)$ at the beginning of the step, then the step leaves the objective 
value unchanged, and yields another optimal solution with $y(M_a)<\mmcap(M_a)$ and 
$y(M_b)<\mmcap(M_b)$.

To summarize, every post-processing step decreases the number of machines $M$ with
$y(M)=\mmcap(M)$.
Hence the entire procedure terminates after at most $m$ steps.
\qed

Let $x^*(J,M)$ and $y^*(M)$ denote an optimal solution of (LP) that satisfies the
property (*) in Lemma~\ref{le:lex.0}. 
Let $\mmm^*$ be the set of machines $M$ with $y^*(M)\ge\mmcap(M)$.
\begin{lemma}
\label{le:lex.1}
The open shop system has a reachable deadlock, if and only if $\mmm^*\ne\emptyset$.
\end{lemma}
\proof
(Only if).
Consider a reachable deadlock state, let $\bbb'$ be the corresponding blocking
set of machines, and let $\jjj'$ be the set of jobs waiting on these machines.
Every job $J\in\jjj'$ is sitting on some machine in $\bbb'$, and is waiting for 
some other machine in $\bbb'$.
Since $|\mmm(J)|=2$, this implies $\mmm(J)\subseteq\bbb'$ for every job $J\in\jjj'$.
Then
\[ 
\sum_{M\in\bbb'} y^*(M) ~\ge~ \sum_{J\in\jjj'} \sum_{M\in\mmm(J)} x^*(J,M) ~=~ |\jjj'|.  
\]
Since furthermore $|\jjj'|=\sum_{M\in\bbb'}\mmcap(M)$, we conclude
$y^*(M)\ge\mmcap(M)$ for at least one machine $M\in\bbb'$.

(If).
Let $\jjj^*$ be the set of jobs with $x^*(J,M)>0$ for some $M\in\mmm^*$.
Property (*) in Lemma~\ref{le:lex.0} now yields the following for every job $J$:
If $J\in\jjj^*$, then $\mmm(J)\subseteq\mmm^*$.
Construct a bipartite graph $G$ between the jobs in $\jjj^*$ and the machines
in $\mmm^*$, with an edge between $J$ and $M$ if and only if $M\in\mmm(J)$.
For any subset $\mmm'\subseteq\mmm^*$, the number of job neighbors in this bipartite
graph is at least $\sum_{M\in\mmm'} y^*(M) \ge \sum_{M\in\mmm'}\mmcap(M)$.
A variant of Hall's theorem from matching theory \cite{LP} now yields that there exists
an assignment of some jobs from $\jjj^*$ to machines in $\mmm^*$ such that every
$M\in\mmm^*$ receives $\mmcap(M)$ pairwise distinct jobs.

To reach a deadlock, we first send all non-assigned jobs one by one through the system.
They are completed and disappear.
Then the assigned jobs enter the system, each moving straightly to the machine to which
it has been assigned.
Then the system falls into a deadlock with blocking set $\mmm^*$: All machines in $\mmm^*$
are full, and all jobs are only waiting for machines in $\mmm^*$.
\qed

Since jobs $J$ with $|\mmm(J)|=1$ are harmless and may be disregarded with respect
to deadlocks, we arrive at the following theorem.

\begin{theorem}
\label{th:lex}
For open shop systems where each job requires processing on at most two machines,
{\probDeadlock} can be solved in polynomial time.  \QED
\end{theorem}

The following example illustrates that the above LP-based approach cannot be carried over
to the case where every job requires processing on three machines (since the only-if part 
of Lemma~\ref{le:lex.1} breaks down).

\begin{example}
Consider a system with two jobs and four machines of unit capacity.
Job $J_1$ needs processing on $M_1,M_2,M_3$, and job $J_2$ needs processing on $M_1,M_2,M_4$.
A (reachable) deadlock results if $J_1$ enters the system on $M_3$ and then moves to $M_1$,
whereas $J_2$ simultaneously enters the system on $M_4$ and then moves to $M_2$.

We consider a feasible solution with $x(J,M)\equiv1/3$ for every $J$ and every 
$M\in\mmm(J)$, and $y(M_1)=y(M_2)=2/3$ and $y(M_3)=y(M_4)=1/3$. 
The ob\-jec\-tive value is~4, and hence this is an optimal solution.
The post-processing leaves the solution untouched, and the resulting set $\mmm^*$ is empty.
\end{example}

\section{Reachable deadlocks if machines have unit capacity}
\label{sec:specialcase.2}
Throughout this section we only consider open shop systems with $\mmcap(M_i)\equiv1$.
For each such system we define a corresponding undirected edge-colored
multi-graph $G=(V,E)$:
The vertices are the machines $M_1,\ldots,M_m$.
Every job $J_j$ induces a clique of edges on the vertex set $\mmm(J_j)$, and all
these edges receive color $c_j$.
Intuitively, if two machines are connected by an edge $e$ of color $c_j$, then
job $J_j$ may move between these machines along edge $e$.

\begin{lemma}
\label{le:unitcap.1}
For an open shop system with unit machine capacities and its corresponding 
edge-colored multi-graph the following two statements are equivalent.
\begin{itemize}
\itemsep=0.0ex
\item[(i)]
The multi-graph contains a simple cycle whose edges have pairwise distinct colors.
\item[(ii)]
The system can reach a deadlock.
\end{itemize}
\end{lemma}
\proof
Assume that (i) holds, and consider a simple cycle $C$ whose edges have pairwise
distinct colors.
By renaming jobs and machines we may assume that the vertices in $C$ are the machines
$M_1,\ldots,M_k$, and that the edges in $C$ are $[M_j,M_{j+1}]$ with color $c_j$
for $1\le j\le k-1$, and $[M_k,M_1]$ with colors $c_k$.
Consider the following processing order of the jobs:
\begin{itemize}
\itemsep=0.0ex
\item
In the first phase, the jobs $J_j$ with $k+1\le j\le n$ are processed one by one:
Job $J_{j+1}$ only enters the system after job $J_j$ has completed all its
processing and has already left the system.
At the end of this phase we are left with the jobs $J_1,\ldots,J_k$.
\item
In the second phase, the jobs $J_1,\ldots,J_k$ are handled one by one.
When job $J_j$ is handled, first all operations of $J_j$ on machines $M_i$ with
$i\ge k+1$ are processed.
Then job $J_j$ moves to machine $M_j$, and stays there till the end of the second phase.
Then the next job is handled.
\end{itemize}
At the end of the second phase, for $1\le i\le k$ job $J_i$ is blocking machine $M_i$,
and waiting for future processing on some other machine in cycle $C$.
The system has fallen into a deadlock, and hence (i) implies (ii).

Next assume that (ii) holds, and consider a deadlock state.
For every waiting job $J_j$ in the deadlock, let $M'_j$ be the machine on which $J_j$
is currently waiting and let $M''_j$ denote one of the machines for which the job is
waiting.
Consider the sub-graph of $G$ that for every waiting job $J_j$ contains the vertex
$M'_j$ together with an edge $[M'_j,M''_j]$ of color $c_j$.
This sub-graph has as many vertices as edges, and hence must contain a simple cycle;
hence (ii) implies (i).
\qed

\begin{lemma}
\label{le:unitcap.2}
For the edge-colored multi-graph $G=(V,E)$ corresponding to some open shop system 
with unit machine capacities, the following three statements are equivalent.
\begin{itemize}
\itemsep=0.0ex
\item[(i)]
The multi-graph contains a simple cycle whose edges have pairwise distinct colors.
\item[(ii)]
The multi-graph contains a 2-vertex-connected component that spans edges of at
least two different colors.
\item[(iii)]
The multi-graph contains a simple cycle whose edges have at least two different colors.
\end{itemize}
\end{lemma}
\proof
We show that (i) implies (ii) implies (iii) implies (i).
The implication from (i) to (ii) is straightforward.

Assume that (ii) holds, and consider a vertex $v$ in such a 2-vertex-connected
component that is incident to two edges with two distinct colors.
These two edges can be connected to a simple cycle, and we get (iii).

Assume (iii), and consider the shortest cycle $C$ whose edges have at least two
different colors.
If two edges $[u,u']$ and $[v,v']$ on $C$ have the same color $c_j$, then
the vertices $u,u',v,v'$ are all in the machine set $\mmm(J_j)$ of job $J_j$.
Hence they span a clique in color $c_j$, and some edges in this clique can be used
to construct a shorter cycle with edges of at least two different colors.
This contradiction shows that (iii) implies (i).
\qed

Lemmas~\ref{le:unitcap.1} and~\ref{le:unitcap.2} together yield that an open shop
system can fall into a deadlock state if and only if the corresponding multi-graph
contains a 2-vertex-connected component that spans edges of at least two different
colors.
Since the 2-vertex-connected components of a graph can easily be determined and
analyzed in linear time (see for instance \cite{CLRS}), we arrive at the following
theorem.
\begin{theorem}
\label{th:unitcap}
For open shop systems with unit machine capacities, problem {\probDeadlock} 
can be solved in polynomial time.  \QED
\end{theorem}



\begin{thebibliography}{8}

\bibitem{BK90}
{\sc Z.A. Banaszak and B.H. Krogh} (1990).
Deadlock avoidance in flexible manufacturing systems with concurrently 
competing process flows.
\emph{IEEE Transactions on Robotics and Automation 6}, 724--734.

\bibitem{CES}
{\sc E.G. Coffman, M.J. Elphick, and A. Shoshani} (1971).
System Deadlocks.
\emph{ACM Computing Surveys 3}, 67--78.

\bibitem{CLRS}
{\sc T.H. Cormen, C.E. Leiserson, R.L. Rivest, and C. Stein} (2001).
\emph{Introduction to Algorithms}.
MIT Press.

\bibitem{GaJo79}
{\sc M.R. Garey and D.S. Johnson} (1979).
\emph{Computers and Intractability: A Guide to the Theory of NP-Completeness}.
Freeman, San Francisco.

\bibitem{Gold78}
{\sc M. Gold} (1978).
Deadlock prediction: Easy and difficult cases.
\emph{SIAM Journal on Computing 7}, 320--336.

\bibitem{LLRS}
{\sc E.L. Lawler, J.K. Lenstra, A.H.G. Rinnooy Kan, and D.B. Shmoys} (1993).
Sequencing and scheduling: Algorithms and complexity.
In: \emph{Handbooks in Operations Research and Management Science, Vol.\ 4},
North Holland, 445--522.

\bibitem{Lawley99}
{\sc M. Lawley} (1999).
Deadlock avoidance for production systems with flexible routing.
\emph{IEEE Transactions on Robotics and Automation 15}, 497--510.

\bibitem{LP}
{\sc L. Lov\'{a}sz and M.D. Plummer} (1986).
Matching Theory.
{\it Annals of Discrete Mathematics 29}, North-Holland.

\bibitem{SuLa01}
{\sc W. Sulistyono and M. Lawley} (2001).
Deadlock avoidance for manufacturing systems with partially ordered process plans.
\emph{IEEE Transactions on Robotics and Automation 17}, 819--832.

\bibitem{XiLiHu01}
{\sc K. Xing, F. Lin, and B. Hu} (2001).
An optimal deadlock avoidance policy for manufacturing system with flexible
operation sequence and flexible routing.
\emph{Proceedings of the 2001 IEEE International Conference on Robotics and
Automation (ICRA'2001)}, 3565--3570.

\end{thebibliography}
\end{document}